\documentclass[12pt]{article}
\usepackage{latexsym}
\usepackage{graphics}
\usepackage{amsmath}
\usepackage{amssymb}

\begin{document}
\centerline{\large \bf Collective effects in traffic on bi-directional ant-trails}

\vspace{2cm}

\centerline{Alexander John{\footnote{e-mail:aj@thp.uni-koeln.de}} and Andreas Schadschneider{\footnote{e-mail:as@thp.uni-koeln.de}}} 
\centerline{Institute for Theoretical Physics, Universit\"at zu K\"oln,}
\centerline{D- 50937 K\"oln, Germany} 
\medskip
\centerline{and}
\medskip
\centerline{Debashish Chowdhury{\footnote{e-mail:debch@iitk.ac.in}}}
\centerline{Department of Physics, Indian Institute of Technology,}
\centerline{Kanpur 208016, India}
\medskip
\centerline{and}
\medskip
\centerline{Katsuhiro Nishinari{\footnote{e-mail:knishi@rins.ryukoku.ac.jp}}}
\centerline{Department of Applied Mathematics and Informatics, Ryukoku University,}
\centerline{Shiga 520-2194, Japan}
\vspace{3cm}

\begin{abstract}
Motivated by recent experimental work of Burd et al., we propose a 
model of bi-directional ant-traffic on pre-existing ant-trails. 
It captures in a simple way some of the generic collective features 
of movements of real ants on a trail. Analyzing this model, we 
demonstrate that there are crucial qualitative differences between 
vehicular- and ant-traffics. In particular, we predict some unusual 
features of the flow rate that can be tested experimentally. As in 
the uni-directional model a non-monotonic density-dependence of the 
average velocity can be observed in certain parameter regimes. As a 
consequence of the interaction between oppositely moving ants the 
flow rate can become approximately constant over some density interval.
\end{abstract}
\vspace{2cm}

\noindent {\bf keywords:}cellular automata, ant trails, collective effects,
computer simulations.

\vspace{2cm}

%
\newpage 
\section{Introduction}
\label{sec1}

At first sight, traffic on ant trails might look rather similar
to human traffic. There are, however, important differences due to
the cooperative nature of ant traffic as we will point out in the
following. An important form of communication between ants is
chemotaxis: Ants deposit pheromones on the substrate as they move 
forward (H\"olldobler and Wilson 1990); other following ants are 
attracted to it and follow the trail. 

Understanding the basic principles governing the formation of the
ant-trails and identifying the factors that influence the movements 
of ants on such trails are of fundamental importance in population
biology of social insect colonies (Wilson 1971). Moreover, insights
gained from these fundamental studies are finding important
applications in computer science (Dorigo et al.1999), communication 
engineering (Bonabeau et al.2000), artificial ``swarm intelligence'' 
(Bonabeau et al.1999) and ``micro-robotics'' (Krieger et al.2000) as 
well as in management (Bonabeau and Meyer 2001).  Furthermore, 
ant-trail is an example of systems of interacting elements driven far 
from equilibrium; the {\it collective spatio-temporal organizations} 
in such systems (Camazine et al.2001, Anderson et al.2002, Mikhailov 
and Calenbuhr 2002, Chowdhury et al.2004) are of current interest in
statistical physics (Schmittmann and Zia 1995, Sch\"utz 2000). Our 
investigation is intended to provide insight into the effects of the 
mutual interactions of the ants on their collective mass flow on the 
trails.

In this paper we focus attention on the ant-traffic on {\em pre-existing} 
ant-trails rather than addressing the question of the formation of 
such trails (Ermentrout and Edelstein-Keshet 1993, Watmough and 
Edelstein-Keshet 1995, Edelstein-Keshet et al.1995, Nicolis and 
Deneubourg 1999, Rauch et al. 1995, Millonas 1992), which is an 
interesting example of self-organization (Camazine et al.2001, 
Anderson et al.2002). In other words, we study ant-traffic on trails 
which persist for very long time because of the availability of 
extensive or renewable resources. 

The similarities between ant-traffic and vehicular traffic
have inspired some recent experimental investigations (Burd et al.
2002, Burd and Aranwela 2003) as well as theoretical modelling 
(Couzin and Franks 2003) of collective movements of ants on trails.  
Our aim is to develop simple models of ant-traffic along the lines 
of discrete models of vehicular traffic (Chowdhury et al.2000, 
Helbing 2001).  Our recent idealized model of uni-directional 
ant-traffic (Chowdhury et al.2002, Nishinari et al.2003) does not 
correspond to the most commonly observed ant-trails. Usually in 
natural trails the traffic is bi-directional with out-bound ants 
proceeding towards the resources to be collected and the nest-bound 
ants carrying the cargo. Therefore, in this paper we propose a new 
model of {\it bi-directional} ant-traffic.

The paper is organized as follows. For the convenience of readers, 
we begin with a brief introduction to the general modelling strategy.
Next, in order to provide insight into the role of pheromone-induced 
indirect interactions among the ants, we briefly review the most 
important unusual features of our earlier model of idealized 
uni-directional ant-traffic (Chowdhury et al.2002, Nishinari et al. 
2003). As we shall show, the uni-directional counterpart helps in 
identifying the physical origins of some of the observed features of 
the bi-directional ant-traffic. Then we introduce our new model of 
bi-directional ant-traffic, present the results of our computer 
simulations and interpret the results physically. In the concluding 
section we summarize our main theoretical predictions and point out 
the current difficulties in comparing these predictions with the 
experimental data available in the literature (Burd et al.2002). 

\section*{\label{sec2}General modeling strategy}

First we describe the general modeling strategy for a uni-directional 
ant-trail model. Later we will generalize this to take into account 
counterflow.

In order to describe the motion of ants one can, in principle, write
differential equations which would be analogs of Newton's equations.
However, in practice, particularly for numerical studies with computer
simulations, it is much simpler to work with {\it discretized} models
(Ermentrout and Edelstein-Keshet 1993) that also take into account 
the self-organized nature of motion.

In such discretized models, continuous space is replaced by a discretized 
mesh of cells; the mesh is usually referred to as a lattice. The size of 
each of these cells is such that it can accomodate at most one ant at a 
time. Therefore, the positions of the ants can change only by discrete 
amounts which can only be integral multiples of the cell size. Depending 
on the ant species under consideration, the typical linear size of these 
cells can be of the order of $0.1$ cm to $1$ cm. 

Moreover, time is also assumed to increase in discrete steps; the 
duration of each step may be taken as the average reflex time of an ant. 
Consequently, the sequence of successive states is like a sequence of 
snapshots of the evolving system.  Since the measured average speeds of 
ants in free-flow (i.e., unhindred by any other ant) has been observed 
(Burd et al.2002) to be of the order of $1$ cm/sec, the real time 
associated with each time step of our model can be of the order of $1$ sec. 
Furthermore, each element of the system (e.g., ant, pheromone, etc) can 
take one of the few allowed discrete states; thus, not only the positions 
but also the velocities of the ants are restricted to a few discrete values. 
In the following we will study the simplest case that ants are
allowed to move to nearest neighbour cells only, i.e.\ the maximal
velocity is 1 cell/time step.

The dynamics of the system, including the movements of the ants, are 
governed by well defined prescriptions that are usually referred to as 
the ``update rules''. Given the state of the system at some arbitrary 
time step $t$, the update rules decide the corresponding state of 
the system at the time step $t+1$. These rules capture the essential 
behavioral features of individual ants, i.e., their responses to their 
immediate local environment (which include their interactions with the 
other neighboring ants). The models with parallel (i.e., synchronous) 
updating rules are usually referred to as cellular automata (CA) 
(Chopard and Droz 1998). The rules need not be deterministic and 
therefore in computer simulations one has to average over different 
realizations of the dynamics. In the context of traffic models, the 
individual-to-individual variations of behavioral patterns are usually 
captured through stochastic rules with appropriate probabilities 
(Chowdhury et al.2000). 

For the sake of simplicity of theoretical calculations, we impose 
periodic boundary conditions which reduces boundary effects. In such 
a discrete model system of $L$ cells with a total of $N$ ants the 
average speed $v$ of the ants is computed as follows: the instantaneous 
average speed $v(t)$ (i.e., speed of the ants {\it averaged over the 
entire population}) is given by $n(t)/N$ where $n(t)$ is the number of 
ants that move forward to the next cell at the time step $t$. However, 
in general, $v(t)$ fluctuates with time; the average of $v(t)$ over 
sufficiently long period of time in the steady state of the system can 
be identified with the average speed $v$. The average number of ants passing 
through a detector location per unit time is called the {\it flux}, 
$F$; it is related to the density $c=N/L$ of the ants and their average 
speed $v$ by the relation $F = c v$. 

All the data reported in this paper have been generated for trails as
large as $L=10^3$, in the units of cell size, i.e., about $1$ to $10$
meters.  In all our computer simulations of the model, we begin with
random initial conditions and let the system evolve following the
update rules specified below. Long after the system reaches its steady
state (typically, $10^4-10^5$ time steps) we begin our computations of
the steady-state properties like, for example, flux, etc. The data are
averaged over the next $10^5-10^6$ time steps as well as sufficient
number of runs, each starting with different initial conditions.

\section*{\label{sec3}Brief review of uni-directional ant-traffic model}

\begin{figure}[tb]
\begin{center}
\resizebox{0.75\textwidth}{!}{\includegraphics{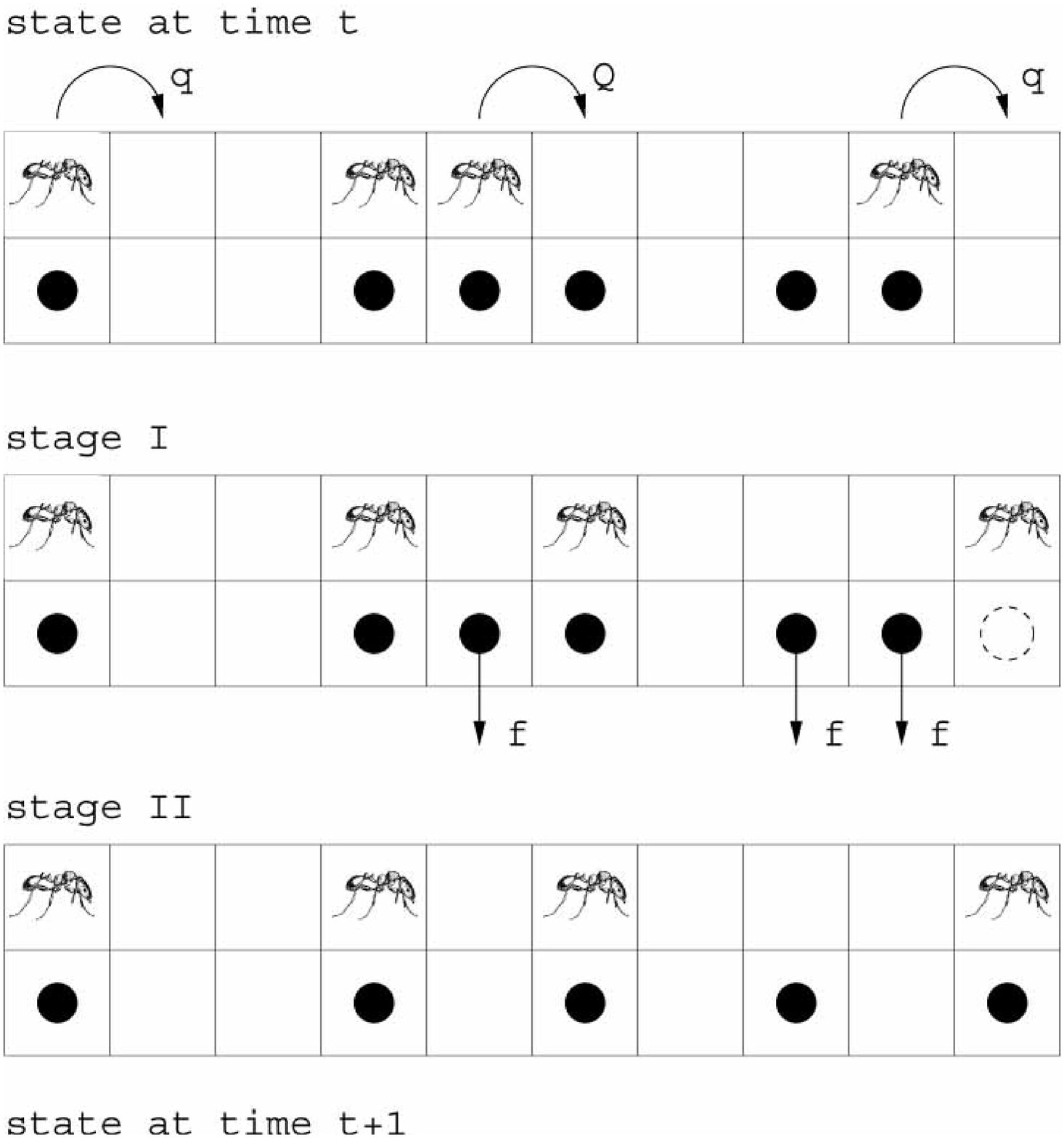}}
\end{center}
\caption{
Schematic representation of typical configurations of the 
{\it uni-directional} ant-traffic model with two different rows of cells
for ants and the pheromones. The symbols $\bullet$ indicate 
the presence of pheromone. 
This figure also illustrates the update procedure. 
Top: Configuration at time $t$, i.e.\ {\it before} stage $I$ 
of the update. The non-vanishing probabilities of forward movement of 
the ants are also shown explicitly. Middle: Configuration {\it after} 
one possible realisation of {\it stage $I$}. Two ants have moved compared 
to the top part of the figure. The open circle with dashed boundary 
indicates the location where pheromone will be dropped by the corresponding 
ant at stage II of the update scheme. Also indicated are the existing 
pheromones, that may evaporate in stage $II$ of the updating, together 
with the average rate of evaporation.  Bottom: Configuration {\it after} 
one possible realization of {\it stage $II$}. Two drops of pheromones 
have evaporated and pheromones has been dropped/reinforced at the 
current locations of the ants.
}
\label{fig-unidir_model}
\end{figure}

In the model of uni-directional circular ant-traffic the trail
consists of one row of cells for the ants and a parallel lattice of
cells for the pheromones (fig.\ref{fig-unidir_model}). Ants are
allowed to move only in one direction (say, clockwise). The state of
the system is updated at {\it each time step} in {\it two stages} (see
fig.\ref{fig-unidir_model}).  In stage I ants are allowed to move
while in stage II the pheromones are allowed to evaporate. In each
stage the {\it stochastic} dynamical rules are applied {\it in
  parallel} to all ants and pheromones, respectively.\\

\noindent {\it Stage I: Motion of ants}\\[0.2cm]
\noindent An ant in a cell cannot move if the cell immediately in front 
of it is also occupied by another ant. However, when an ant finds that  
the cell immediately in front of it is not occupied by any other ant, 
the likelihood of its forward movement to the ant-free cell is $Q$ or $q$, 
per unit time step, depending on whether or not the target cell 
contains pheromone. Thus, $q$ (or $Q$) would be the average speed of a 
{\it free} ant in the absence (or presence) of pheromone. To be consistent 
with real ant-trails, we assume $ q < Q$, as presence of pheromone 
increases the average speed.\\

\noindent {\it Stage II: Evaporation of pheromones}\\[0.2cm]
\noindent Trail pheromone is volatile. So, the deposited pheromone 
will gradually decay unless reinforced by the following ants. In order to 
capture this process, we assume that each cell occupied by an ant at the 
end of stage I also contains pheromone. On the other hand, pheromone in 
any `ant-free' cell is allowed to evaporate; this evaporation is also 
assumed to be a random process that takes place at an average rate of $f$ 
per unit time. 

For the sake of simplicity, we consider only the presence or absence
of pheromone in a cell. However, it will be straightforward to
generalize the model to describe different levels of strength of the
pheromone as well as their gradual evaporation and even diffusion.
Moreover, in real ant trails, the extent of reinforcement of the trail
by dropping of pheromone may depend on the existing local strength of
pheromone. But, these details are not incorporated in our model since
we do not explicitly treat the different levels of strength of the
pheromone. Besides, incorporation of these details is not expected to
affect the qualitative features of the results.  The total amount of
pheromone on the trail can fluctuate although the total number of the
ants is independent of time because of the periodic boundary
conditions.

\begin{figure}[tb]
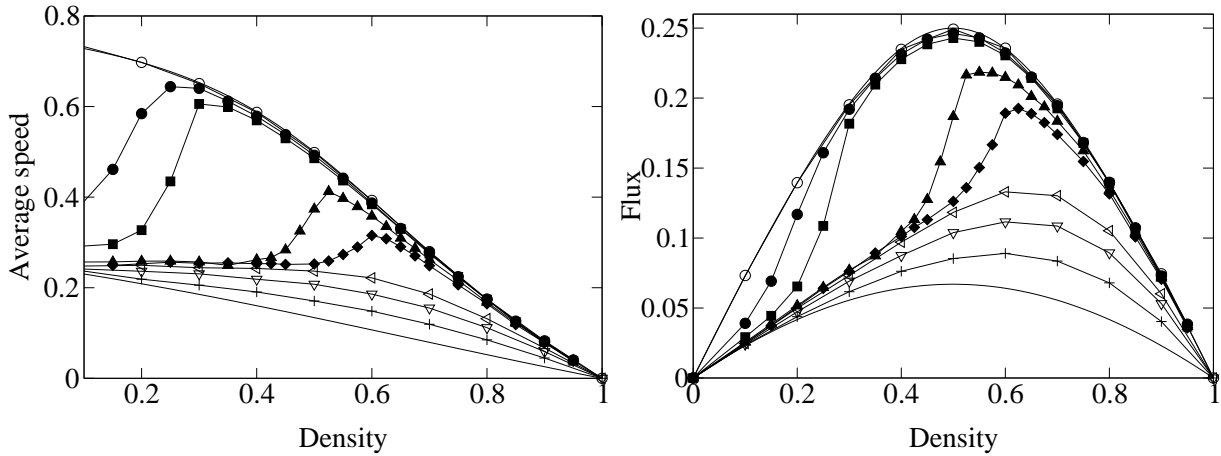

\begin{center}
\vspace{0.5cm}
\resizebox{0.47\textwidth}{!}{\includegraphics{uni_speed.eps}}
\resizebox{0.47\textwidth}{!}{\includegraphics{uni_flux.eps}}
\end{center}
\caption{ 
The average speed (left) and the flux (right) of the ants, in the 
{\it uni-directional} ant-traffic model, are plotted against
their densities for the parameters $Q = 0.75, q = 0.25$.
The discrete data points corresponding to $f=0.0001 ({\circ})$,
$0.0005 (\bullet$), $0.001 (\blacksquare)$, $0.005 (\blacktriangle)$ 
$0.01 ({\blacklozenge})$, $0.05 ({\triangleleft})$, 
$0.10 (\bigtriangledown)$, $0.25 ($+$)$ have been obtained from 
computer simulations; the lines connecting these data points merely 
serve as the guide to the eye. The cases $f=0$ and 
$f=1$ are also displayed by the uppermost and lowermost curves 
(without points); these are exact results. Curves plotted with 
filled symbols have unusual shapes.}
\label{fig-unidir_res}
\end{figure}

In the two special cases $f = 0$ and $f = 1$ this model of 
uni-directional ant traffic becomes identical to one of the special 
cases of the Nagel-Schreckenberg model (Nagel and Schreckenberg 1992) 
of vehicular traffic with the corresponding likelihoods of forward 
movements $Q$ and $q$, respectively. 
In vehicular traffic, usually, the inter-vehicle interactions tend to 
hinder each other's motion so that the {\it average speed} of the 
vehicles decreases {\it monotonically} with the increasing density of 
the vehicles. In contrast, in our model of uni-directional ant-traffic 
the average speed of the ants varies {\it non-monotonically} with their 
density over a wide range of small values of $f$ 
(see fig.\ref{fig-unidir_res}, left) because of the coupling of their 
dynamics with that of the pheromone. This uncommon variation of the 
average speed gives rise to the unusual dependence of the flux on the 
density of the ants in our uni-directional ant-traffic model 
(fig.\ref{fig-unidir_res}, right).

\section*{\label{sec4}The model of bi-directional ant-traffic}

We develope the model of bi-directional ant-traffic by extending the 
model of uni-directional ant-traffic described in the previous
section.

\subsection*{The model}

\begin{figure}[tb]
\begin{center}
\resizebox{0.5\textwidth}{!}{\includegraphics{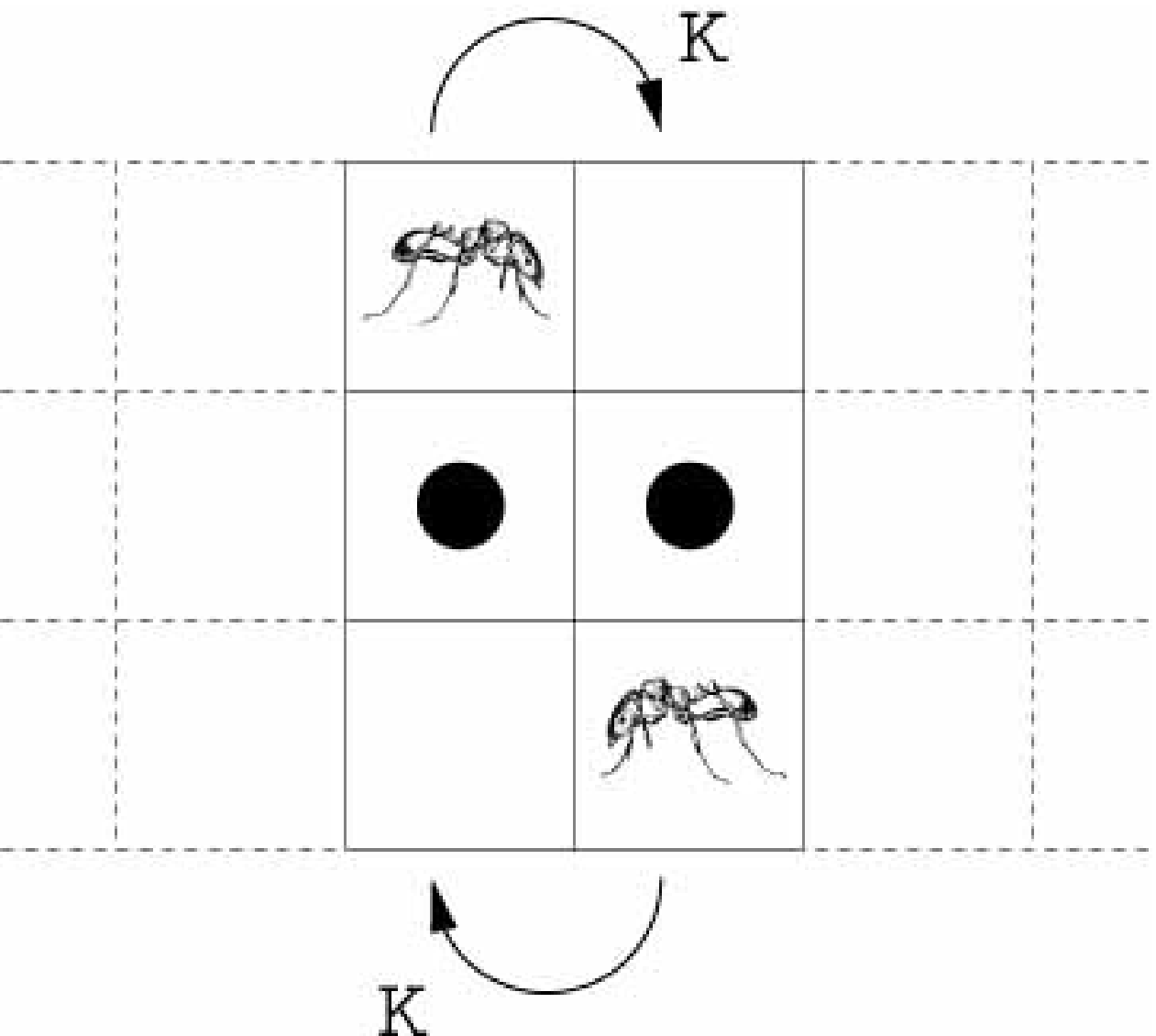}}
\end{center}
\caption{A typical head-on encounter of two ants moving in opposite 
directions in our model of {\it bi-directional} ant-traffic.
This is a totally new process which does not have any analog in the 
model of uni-directional ant-traffic. The mesh of cells in the
middle corresponds to the common pheromone trail.  
}
\label{fig-bi}
\end{figure}

In the models of bi-directional ant-traffic the trail consists of {\it
  two} lanes of cells for the ants (see fig.\ref{fig-bi}). These two
lanes need not to correspond to physically separate rigid lanes in
real space; these are, however, convenient for describing the
movements of ants in two opposite directions. In the initial
configuration, a randomly selected subset of the ants move in the
clockwise direction in one lane while the others move counterclockwise
in the other lane. However, ants are allowed neither to take U-turn
\footnote{Although U-turn of foragers is not uncommon (Beckers et al.
1992), U-turn of followers on pre-existing trails is very rare.} nor 
to change lane. Thus, the ratio of the populations of clockwise-moving
and anti-clockwise moving ants remains unchanged as the system evolves
with time. All the numerical data presented in this paper have been
obtained by Monte Carlo simulations of the {\it symmetric} case where 
an equal number of ants move in the two directions. Therefore, the 
{\it average} flux is identical for both directions. In all the graphs 
we plot only the flux for clockwise-moving ants.

The rules governing the deposition and evaporation of pheromone in the
model of bi-directional ant-traffic are identical to those in the
model of uni-directional traffic. The {\it common} pheromone trail is
created and reinforced by both the clockwise and counterclockwise
moving ants. The probabilities of forward movement of the ants in the
model of bi-directional ant-traffic are also natural extensions of the
similar situations in the uni-directional traffic.  When an ant (in
either of the two lanes) {\it does not} face any other ant approaching
it from the opposite direction the likelihood of its forward movement
onto the ant-free cell immediately in front of it is $Q$ or $q$,
respectively, depending on whether or not it finds pheromone ahead.
Finally, if an ant finds another oncoming ant just in front of it, as
shown in fig.\ref{fig-bi}, it moves forward onto the next cell with
probability $K$; in such situations, pheromone is present in both the 
cells.

Since in reality ants do not segregate in perfectly well defined lanes, 
head-on encounters of oppositely moving individuals occur quite often 
although the frequency of such encounters and the lane discipline 
varies from one species of ants to another (Burd et al.2002). In reality, 
two ants approaching each other feel the hindrance, turn by a small 
angle to avoid head-on collision (Couzin and Franks 2003) and,
eventually, pass each other.  At first sight, it may appear that the
ants in our model follow perfect lane discipline and, hence, the model
dynamics is unrealistic. However, that is not true.  The violation of
lane discipline and head-on encounters of oppositely moving ants is
captured, effectively, in an indirect manner by assuming $K < Q$. 
Since, as in the uni-directional case, we have $q< Q$ this implies the 
existence of two different parameter regimes, $q< K <Q$ and $K<q<Q$.
It is worth mentioning here that even in the special limit $K =
Q$ the traffic dynamics on two lanes would remain coupled because the
pheromone dropped by the ants on one lane also influences the ants 
moving on the other lane in the opposite direction.

\subsection*{Results and interpretations}

Figs.~\ref{fig-bifd1} and \ref{fig-bifd2} show the variations of flux 
and average speeds with density of ants in our model for the two 
relevant cases $q<K<Q$ and $K<q<Q$ and different values of the 
evaporation probability $f$. In both cases, the non-monotonic variation 
of the average speed with density gives rise to the unusual shape of 
the flux-versus-density diagram over a range of values of $f$. This 
feature of the model of bi-directional traffic is similar to that of 
the uni-directional ant-traffic (compare the figs.\ref{fig-bifd1} 
and \ref{fig-bifd2} with fig.\ref{fig-unidir_res}). It results from 
the formation of a `loose cluster' (Nishinari et al.2003), i.e.\ a
localized region of increased density where the ants move almost
coherently.

Another interesting phenomenon related to the cluster formation is
coarsening.  At intermediate time usually several loose clusters are
formed.  However, the velocity of a cluster depends on the distance to
the next cluster ahead. This velocity is controlled by the survival 
probability $p_s$ of the pheromone created by the last ant of the 
previous cluster. Obviously, $p_s$ decreases with increasing distance. 
Therefore clusters with a small headway move faster than those with 
a large headway.  This induces a coarsening process such that after 
long times only one loose cluster survives.

For both parameter regimes, the influence of ants in the
counterdirection is determined by the evaporation probability $f$. In
the regime $q<K<Q$ this influence is not necessarily repulsive.  For
$f\ll 1$ the ants move predominantly with the hopping probability $K$
or $Q$. The hopping probability is decreased by ants moving in the
opposite direction and therefore the coupling is repulsive.  But at
high evaporation rates $f\approx 1$ the coupling becomes attractive
for $q<K$.  In the regime $K<q<Q$ one finds in principle the same
mechanism. For high evaporation rates the coupling is repulsive
since $K<q$ and becomes even more repulsive as the evaporation probability
decreases such that more ants move with probability $Q$.

\begin{figure}[tb]
\begin{center}
\vspace{0.5cm}
\resizebox{0.47\textwidth}{!}{\includegraphics{johnspeed1.eps}}
\resizebox{0.47\textwidth}{!}{\includegraphics{johnflux1.eps}}
\end{center}
\caption{The average speed (left) and the flux (right) of the ants in the 
model of {\it bi-directional} traffic are plotted as functions of 
their density for the case $q<K<Q$ and several different values of 
the pheromone evaporation probability $f$. 
The parameters are $Q=0.75, q = 0.25$ 
and $K=0.5$. The symbols  $\circ$, $\bullet$, $\blacksquare$, 
$\bigtriangleup$, $\ast$, $+$, $\bigtriangledown$, $\Diamond$  
and $\triangleleft$ correspond, respectively, to 
$f = 0, 0.0005, 0.005,0.05, 0.075,0.10,0.25,0.5$ and $1$.  
The lines are merely guides to the eye.
}
\label{fig-bifd1}
\end{figure}


\begin{figure}[tb]
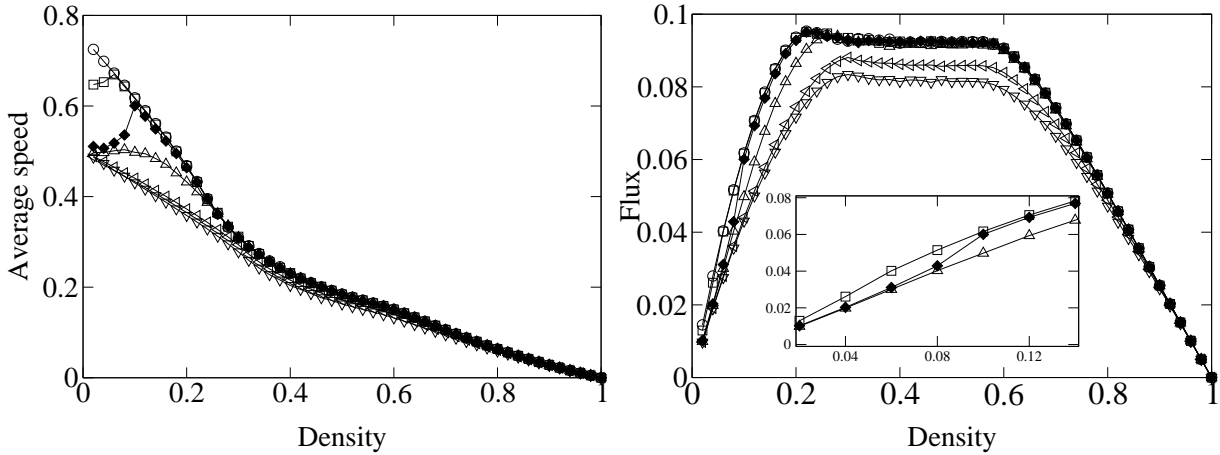

\begin{center}
\vspace{0.5cm}
\resizebox{0.47\textwidth}{!}{\includegraphics{johnspeed2.eps}}
\resizebox{0.47\textwidth}{!}{\includegraphics{johnflux2.eps}}
\end{center}
\caption{The average speed (left) and the flux (right) of the ants in the 
model of {\it bi-directional} ant traffic are plotted as functions of 
their density for the case $K<q<Q$ and several different values of 
the pheromone evaporation probability $f$. 
The parameters are $Q=0.75, q = 0.50$ and $K=0.25$.
The symbols $\circ$, $\square$, $\blacklozenge$, $\bigtriangleup$, 
$\triangleleft$ and $\bigtriangledown$ correspond, respectively, to 
$f = 0, 0.0005, 0.005, 0.05, 0.5$ and $1$.  
The lines are merely guides to the eye. The inset in the left diagram
is a magnified re-plot of the same data, over a narrow range of density, 
to emphasize the fact that the unusual trend of variation of flux with 
density in this case is similar to that observed in the fig.\ref{fig-bifd1}.
}
\label{fig-bifd2}
\end{figure}
\newpage

An additional feature of the density-dependence of the flux in the
bi-directional ant-traffic model is the occurrence of a plateau
region.  This plateau is more pronounced for $K<q<Q$
(fig.\ref{fig-bifd2}) than for $q<K<Q$ (fig.\ref{fig-bifd1}). Such
plateaus in the flux-versus-density diagram have been observed earlier
(Janowsky and Lebowitz 1992, Tripathy and Barma 1997) in models related 
to vehicular traffic where randomly placed bottlenecks ('defects') 
slow down the traffic in certain locations along the route. Here the 
mechanism is similar although in principle we have dynamical defects.
Note that in fig.\ref{fig-bifd1} the plateaus appear only in the two 
limits $f \rightarrow 0$ and $f \rightarrow 1$ but not for an 
intermediate range of values of $f$. In the limit $f \rightarrow 0$, 
most often the likelihood of the forward movement of the ants is 
$Q = 0.75$ whereas they are forced to move with a smaller probability 
$K = 0.5$ at those locations where they face another ant immediately 
in front approaching from the opposite direction (like the situations 
depicted in fig.\ref{fig-bi}). Thus, such encounters of oppositely 
moving ants have the same effect on ant-traffic as that of the 
bottlenecks on vehicular traffic.

But why do the plateaus re-appear in fig.\ref{fig-bifd1} also 
in the limit $f \rightarrow 1$? At sufficiently high densities, 
oppositely moving ants facing each other move with probability $K = 0.5$ 
rather than $q = 0.25$. In this case, locations where the ants have to 
move with the lower probability $q$ will be, effectively bottlenecks and 
hence the re-appearance of the plateau. As $f$ approaches unity there 
will be larger number of such locations and, hence, the wider will be the 
plateau. This is consistent with our observation in fig.\ref{fig-bifd1}.

\section*{\label{sec5} Summary}

In this paper we have introduced a model of {\it bi-directional} 
ant-traffic. The two main theoretical predictions of this model 
are as follows:\\
(i) The average speed of the ants varies {\it non-monotonically} 
with their density over a wide range of pheromone evaporation rates. 
This unusual variation of average speed with density gives rise 
to the uncommon shape of the flux-versus-density diagrams 
and has already been observed in the uni-directional model.\\
(ii) Over some regions of parameter space, the flux exhibits 
plateaus when plotted against density. This new feature is
characteristic for the bi-directional model and has its origin
in the mutual hindrance of ants moving in opposite directions.

In principle, it should be possible to test these theoretical 
predictions by comparing with the corresponding experimental data.
Interestingly, various aspects of locomotion of individual ants have 
been studied in quite great detail (Lighton et al.1987, Zollikofer 
1994, Weier et al. 1995). However, traffic is a collective phenomenon 
involving a large number of interacting ants. Surprisingly, to our 
knowledge, the results published by Burd et al.(2002) on the 
leaf-cutting ant {\em Atta Cephalotes} are the only set of 
experimental data available on the density-dependence of flux of 
ants on trails. Unfortunately, the fluctuations in the data 
are too high to make any direct comparison with our theoretical 
predictions.

We hope our predictions will motivate more  experimental 
measurements. The new experiments should be carried out, preferably, 
on a {\it circular} trail to mimic the periodic boundary conditions. 
Since we assumed all the ants to have identical size, the experiments 
should be done with only one type of ants, although polymorphism
is quite common. Furthermore, in order to see the effect of $f$ on 
the flux, the experiments should be repeated with different species 
of ants whose trail pheromones evaporate at significantly different 
rates. The typical magnitudes of $f$, for which the non-monotonic 
variation of the average speed with density is predicted, correspond 
to pheromone lifetimes in the range from few minutes to tens of minutes.

In the models developed so far we do not distinguish between the 
characteristic features of outbound and nest-bound ants although, 
small variations in their free-flow velocities  is expected in 
real ant-trails (Burd et al.2002). We are now investigating the effects 
of this difference in the average speeds of nest-bound and outbound 
ants as well as that of open boundary conditiond on the flux. The 
results of these studies, together with the results for unequal 
populations of the outbound and nestbound ants, will be reported in 
detail elsewhere (John et al.2004).

For a direct application to ant trails the use of open boundary
conditions (with `reservoirs' representing e.g.\ nest and food source) 
would be more realistic. Investigations of this situation are currently 
being carried out (John et al.2004). However, from previous experiences 
with nonequilibrium systems one expects that the behaviour of the open 
system, e.g.\ the phase diagram, will be determined by that of the 
periodic system studied here (Sch\"utz 2000).  \\[0.5cm]

{\bf Acknowledgments}\\[0.2cm]
We thank Martin Burd for an enlightening discussion, Iain Couzin for 
useful correspondence and Bert H\"olldobler for drawing our attention 
to the paper by Burd et al.(2002). One of the authors (DC) acknowledges 
support, in part, from the Deutsche Forschungsgemeinschaft (DFG) through 
a joint Indo-German joint research project. 

\vspace{0.9cm}


{\bf Bibliography}

\noindent Anderson C, Theraulaz G, and Deneubourg J L, 2002.
{\it Self-assemblages in insect societies}, 
Insectes Soc. {\bf 49}, 99-110.\\

\noindent Beckers R, Deneubourg J L, and Goss S, 1992.
{\it Trails and U-turns in the selection of a path by the ant {\em Lasius niger}},
J. Theor. Biol. {\bf 159},  397-415.\\

\noindent Bonabeau E, Dorigo M, and Theraulaz G, 2000.
{\it Inspiration for optimization from social insect behaviour},
Nature {\bf 400}, 39-42. \\  

\noindent Bonabeau E, Dorigo M, and Theraulaz G, 1999.
{\it Swarm Intelligence: From Natural to Artificial Systems},
(Oxford University Press).\\

\noindent Bonabeau E and Meyer C, 2001.  {\it Swarm intelligence: 
A whole new way to think about business}, 
Harvard Business Review, May issue, 107-114.\\

\noindent Burd M, Archer D, Aranwela N, and Stradling D J, 2002.
{\it Traffic dynamics of the leaf-cutting ant, {\em Atta cephalotes}}, 
American Natur. {\bf 159}, 283-293.\\

\noindent Burd M and Aranwela N, 2003.
{\it Head-on encounter rates and walking speed of foragers in leaf-cutting 
ant traffic}, 
Insectes Soc. {\bf 50}, 3-8.\\

\noindent Camazine S, Deneubourg J L, Franks N R, Sneyd J, Theraulaz G, 
and Bonabeau E, 2001. {\it Self-organization in Biological Systems} 
(Princeton University Press).\\

\noindent Chopard B and Droz M, 1998.
{\it Cellular Automata Modelling of Physical Systems},
(Cambridge University Press).\\

\noindent Chowdhury D, Santen L and Schadschneider A, 2000.
{\it Statistical physics of vehicular traffic and some related systems}, 
Phys. Rep. {\bf 329}, 199-329.\\

\noindent Chowdhury D, Guttal V, Nishinari K, and Schadschneider A, 2002. 
{\it A cellular-automata model of flow in ant-trails: Non-monotonic variation 
of speed with density}, 
J. Phys. A: Math. Gen. {\bf 35}, L573-L577.\\

\noindent Chowdhury D, Nishinari K, and Schadschneider A, 2004.
{\it Self-organized patterns and traffic flow in colonies of organisms: from bacteria and social insects to vertebrates},
to appear in Phase Trans. (e-print {\tt q-bio.PE/0401006}).\\

\noindent Couzin I D and Franks N R, 2003.
{\it Self-organized lane formation and optimized traffic flow in army ants}, 
Proc. Roy. Soc. London B {\bf 270}, 139-146.\\

\noindent Dorigo M, di Caro G, and  Gambardella L M, 1999.
{\it Ant algorithms for discrete optimization},
Artificial Life  {\bf 5}(3), 137-172.\\

\noindent Edelstein-Keshet L, Watmough J, and Ermentrout G B, 1995.
{\it Trail-following in ants: individual properties determine population behaviour}, 
Behav. Ecol. and Sociobiol. {\bf 36}, 119-133.\\

\noindent Ermentrout G B and Edelstein-Keshet L, 1993.
{\it Cellular automata approaches to biological modeling}, 
J. Theor. Bio. {\bf 160}, 97-133. \\ 

\noindent Helbing D, 2001. 
{\it Traffic and related self-driven many-particle systems}, 
Rev. Mod. Phys. {\bf 73}, 1067-1141.\\

\noindent H\"olldobler B and  Wilson E O, 1990. {\it The ants} 
(The Belknap Press of Harvard University Press, Cambridge, Mass., USA). \\

\noindent Janowsky S A and Lebowitz J L, 1992.
{\it Finite-size effects and shock fluctuations in the asymmetric 
simple-exclusion process}, Phys. Rev. A {\bf 45}, 618-625.\\ 

\noindent John A, Kunwar A, Chowdhury D, Nishinari K, and Schadschneider A,
2004. (to be published).\\

\noindent Krieger, M, Billeter J B, and Keller L, 2000.
{\it Ant-like task allocation and recruitment in cooperative robots},
Nature {\bf 406}, 992-995.\\

\noindent Lighton J R B, Bartholomew G A, and Feener D H, 1987.
{\it Energetics of locomotion and load carriage and a model of the energy 
cost of foraging in the leaf-cutting ant {\it Atta colombica}}, 
Physiological Zoology {\bf 60}, 524-537.\\

\noindent Mikhailov A S and Calenbuhr V, 2002.
{\em From Cells to Societies} (Springer, Berlin).\\

\noindent Millonas M M, 1992.
{\it A connectionist type model of self-organized foraging and emergent 
behavior in ant swarms}, 
J. Theor. Biol. {\bf 159}, 529-552.\\

\noindent Nagel K and Schreckenberg M, 1992.
{\it A cellular automaton model for freeway traffic},
J. Physique I {\bf 2}, 2221-2229.\\

\noindent Nicolis S C and  Deneubourg J L, 1999.
{\it Emerging patterns and food recruitment in ants: an analytical study}, 
J. Theor. Biol. {\bf 198}, 575-592.\\

\noindent Nishinari K, Chowdhury D, and Schadschneider A, 2003.
{\it Cluster formation and anomalous fundamental diagram in an ant trail model},
Phys. Rev. E {\bf 67}, 036120.\\

\noindent Rauch E M, Millonas M M, and Chialvo D R, 1995.
{\it Pattern formation and functionality in swarm models}, 
Phys. Lett. A {\bf 207}, 185-193.\\

\noindent Schmittmann B and Zia R.K.P., 1995. 
{\it Statistical Mechanics of Driven Diffusive Systems}, 
In: {\it Phase Transitions and Critical Phenomena, Vol.17} 
(C. Domb and J.L. Lebowitz (Eds.), Academic Press).\\

\noindent Sch\"utz G, 2000.
{\it Exactly Solvable Models For Many-Body Systems Far From Equilibrium}, 
In:  {\it Phase Transitions and Critical Phenomena, Vol.19} 
(C. Domb and J.L. Lebowitz (Eds.), (Academic Press).\\

\noindent Tripathy G and Barma M, 1997.
{\it Steady state and dynamics of driven diffusive systems with quenched disorder}, Phys. Rev. Lett. {\bf 78}, 3039-3042.\\

\noindent Watmough J and Edelstein-Keshet L, 1995. 
{\it Modelling the formation of trail networks by foraging ants}, 
J. Theor. Biol.~{\bf 176}, 357-371.\\

\noindent Weier J A, Feener D H, and Lighton J R B, 1995.
{\it Inter-individual variation in energy cost of running and loading in the 
seed-harvester ant {\it Pogonomyrmex maricopa}}, 
J. Insect Physiology {\bf 41}, 321-327.\\

\noindent Wilson E O, 1971. {\it The Insect Societies} (The Belknap Press of
Harvard University Press, Cambridge, Mass., USA).\\

\noindent Zollikofer C P E, 1994.
{\it Stepping patterns in ants. 1, 2, 3}, 
J. Experimental Biology {\bf 192}, 95-127.\\

\end{document}